\documentclass[11pt]{article}
\usepackage[top=1.5in,bottom=1.5in,left=1.5in,right=1.5in]{geometry}

\usepackage{color}
\usepackage{graphicx}
\usepackage{amsmath}
\usepackage{amssymb}
\usepackage{amsthm}
\usepackage{verbatim}
\usepackage{listings}
\usepackage{subfigure}
\usepackage[authoryear]{natbib}
\newcommand{\iid}{\stackrel{\emph{i.i.d.}}{\sim}}
\newcommand{\DF}{\text{DF}}
\newcommand{\bs}{\boldsymbol}
\newcommand{\tr}{\ensuremath{\operatorname{tr}}}
\newcommand{\var}{\ensuremath{\operatorname{Var}}}
\newcommand{\cov}{\ensuremath{\operatorname{Cov}}}
\newtheorem{1}{Theorem}
\newtheorem{2}{Remark}
\newtheorem{3}{Definition}
\newtheorem{4}[3]{Definition}
\newtheorem{5}[1]{Theorem}

\DeclareMathOperator*{\argmin}{arg\,min}

\begin{document}


\title{Effective Degrees of Freedom: A Flawed Metaphor}

\author{Lucas Janson, Will Fithian, Trevor Hastie}


\maketitle

\begin{abstract}
To most applied statisticians, a fitting procedure's degrees of
freedom is synonymous with its model complexity, or its capacity for
overfitting to data.  In particular, it is
often used to parameterize the bias-variance tradeoff in model
selection. We argue that, contrary to folk intuition, model complexity
and degrees of freedom are {\em not} synonymous and may correspond
very poorly.  We exhibit and theoretically explore various
examples of fitting procedures for which degrees of freedom is not
monotonic in the model complexity parameter, and can exceed the total
dimension of the response space.  Even in very simple settings, the
degrees of freedom can exceed the dimension of the ambient space by an
arbitrarily large amount. We show the degrees of freedom for any non-convex projection method can be unbounded.
\end{abstract}




\section{Introduction}\label{Intro}

Consider observing data $\boldsymbol{y} = \bs{\mu} + \boldsymbol{\varepsilon}$ with fixed mean $\bs{\mu}\in \mathbb{R}^n$ and mean-zero errors $\boldsymbol{\varepsilon} \in \mathbb{R}^n$, and predicting $\bs{y}^*=\bs{\mu} + \bs{\varepsilon}^*$, where $\bs{\varepsilon}^*$ is an independent copy of $\bs{\varepsilon}$. Assume for simplicity the entries of $\bs{\varepsilon}^*$ and $\bs{\varepsilon}$ are independent and identically distributed with variance $\sigma^2$.

Statisticians have proposed an immense variety of fitting procedures for producing the prediction $\hat{\bs{y}}(\bs{y})$, some of which are more complex than others. The effective degrees of freedom (DF) of \cite{Efron1986}, defined precisely as $\frac{1}{\sigma^2} \sum^n_{i = 1} \cov\big(y_i, \hat{y}_i\big)$, has emerged as a popular and convenient measuring stick for comparing the complexity of very different fitting procedures (we motivate this definition in Section~\ref{Prelim}). The name suggests that a method with $p$ degrees of freedom is similarly complex as linear regression on $p$ predictor variables.

To motivate our inquiry, consider estimating a no-intercept
linear regression model with design matrix
$\boldsymbol{X}=\begin{pmatrix} 1 & 0\\ 0 & 1\end{pmatrix}$ and
response $\bs{y} \sim N(\bs{\mu}, I)$, with $\bs{\mu} \in
\mathbb{R}^2$.  Suppose further that, in order
to obtain a more parsimonious model than the full bivariate
regression, we instead estimate the best fitting of the two univariate
models (in other words, best subsets regression with model size
$k=1$). In effect we are minimizing training error over the model $ \bs{\mu} \in \mathcal{M} =\mathbb{R} \times \{0\} \cup \{0\} \times \mathbb{R}$.
What are the effective degrees of freedom of the fitted model in
this seemingly innocuous setting?

A simple, intuitive, and wrong argument predicts
that the DF lies somewhere between 1 and 2.  We expect
it to be greater than 1, since we use the data to select the better of two
one-dimensional models.  However, we have only two free parameters at
our disposal, with a rather severe constraint on the values they
are allowed to take, so $\mathcal{M}$ is strictly less complex than
the saturated model with two parameters that fits the data as hard as
possible.

\begin{figure}[ht!]
  \centering
  \includegraphics[width=.65\textwidth]{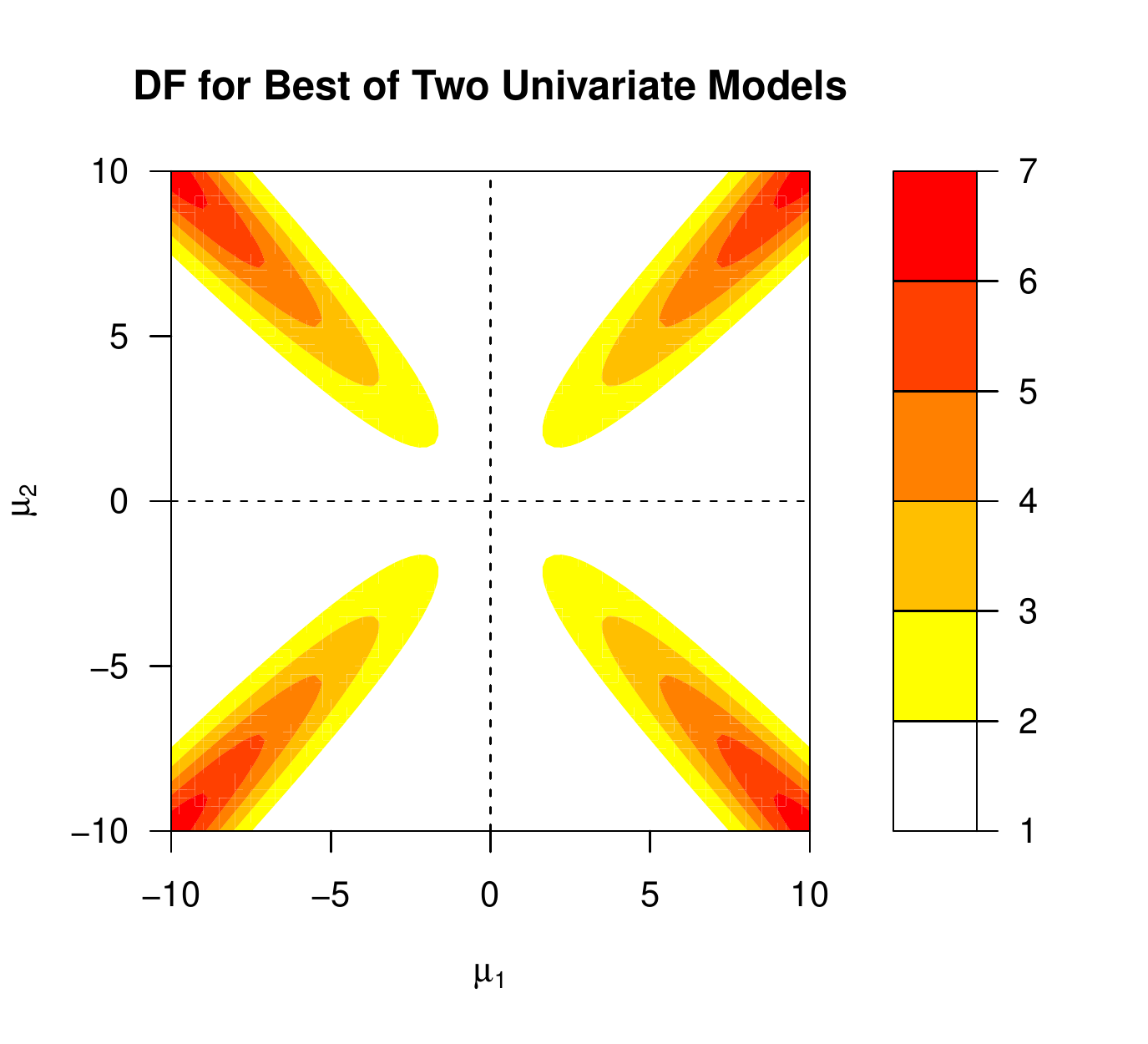}
  \caption{Heatmap of the DF for 1-best-subset fit with the model ${\bs{y}
    \sim N(I_2 \cdot \bs{\beta}, I_2\cdot \sigma^2)}$, as a function of the true mean
    vector $\bs{\mu} \in \mathbb{R}^2$.  Contrary to what one
    might naively expect, the DF can significantly exceed 2, the DF
    for the full model.}
  \label{fig:dfmap}
\end{figure}

Figure~\ref{fig:dfmap} shows the effective DF for this
model, plotted as a function of $\boldsymbol{\mu}\in\mathbb R^2$, the expectation of
the response $\boldsymbol{y}$.  As expected, $\DF(\boldsymbol{\mu})\geq 1$, with
near-equality when $\boldsymbol{\mu}$ is close to one of the coordinate axes but far from
the other.  Perhaps surprisingly, however, $\DF(\boldsymbol{\mu})$ can exceed 2,
approaching 7 in the corners of the plot.  Indeed, the plot suggests
(and we will later confirm)
that $\DF(\boldsymbol{\mu})$ can grow arbitrarily large as
$\boldsymbol{\mu}$ moves farther diagonally from the origin.

To understand why our intuition should lead us astray here, we must
first review how the DF is defined for a general fitting procedure,
and what classical concept that definition is meant to generalize.

\subsection{Degrees of Freedom in Classical Statistics}

The original meaning of degrees of freedom,
the number of dimensions in which a random vector may vary, plays a
central role in classical statistics.
In ordinary linear regression with full-rank $n\times p$ predictor matrix $\boldsymbol{X}$, the fitted response $\hat{\boldsymbol{y}} =
\boldsymbol{X}\hat{\boldsymbol{\beta}}$ is the orthogonal projection of
$\boldsymbol{y}$ onto the $p$-dimensional column space of
$\boldsymbol{X}$, and the residual
$\boldsymbol{r}=\boldsymbol{y}-\hat{\boldsymbol{y}}$ is the projection
onto its orthogonal complement, whose
dimension is $n-p$.  We say this linear model has $p$
``model degrees of freedom'' (or just ``degrees of freedom''), with
$n-p$ ``residual degrees of freedom.''

If the error variance is $\sigma^2$, then
$\boldsymbol r$ is ``pinned down'' to have zero projection in $p$
directions, and is free to vary, with variance $\sigma^2$,
in the remaining $n-p$ orthogonal directions.  In particular, if the
model is correct ($\mathbb E [\boldsymbol{y}] =
\boldsymbol{X}\boldsymbol{\beta}$), then the residual sum of squares (RSS)
has distribution
\begin{equation}
  \text{RSS} = \|\boldsymbol r\|_2^2 \sim \sigma^2 \cdot \chi_{n-p}^2.
\end{equation}
It follows that $\mathbb{E} \left[\|\boldsymbol r\|_2^2\right] = \sigma^2(n-p)$,
leading to the unbiased variance estimate $\hat\sigma^2 =
\frac{1}{n-p}\|\boldsymbol r\|_2^2$. $t$-tests and
$F$-tests are based on comparing lengths of $n$-variate
Gaussian random vectors after projecting onto appropriate linear subspaces.

In linear regression, the
model degrees of freedom (henceforth DF) serves to quantify multiple
related properties of the fitting procedure.
The DF coincides with the
number of non-redundant free parameters in the model, and thus
constitutes a natural measure of model complexity or overfitting.
In addition, the
total variance of the fitted response $\hat{\boldsymbol{y}}$ is
exactly $\sigma^2p$, which depends only on the number of linearly
independent predictors and not on their size or correlation with each
other.

The DF also quantifies {\em optimism} of the residual sum of squares
as an estimate of out-of-sample prediction error.  In linear regression,
one can easily show that the RSS understates mean squared
prediction error by $2\sigma^2p$ on average.
\cite{Mallows1973} proposed exploiting this identity as a means of
model selection, by computing ${\text{RSS} + 2\sigma^2}p$, an unbiased estimate
of prediction error, for several models, and selecting the model with
the smallest estimated test error.  Thus, the DF of each model
contributes a sort of penalty for how hard that model is fitted to the
data.

\subsection{``Effective'' or ``Equivalent'' Degrees of Freedom}

For more general fitting procedures such as smoothing splines,
generalized additive models, or ridge regression, the number
of free parameters is often either undefined or an inappropriate measure of
model complexity.  Most such methods feature some tuning parameter
modulating the fit's complexity, but it is not clear a priori how to
compare e.g. a Lasso with Lagrange parameter $\lambda=3$ to a local
regression with window
width $0.5$.  When comparing different methods, or the same method
with different tuning parameters, it can be quite useful to have some
measure of complexity with a consistent meaning across a diverse range
of algorithms.  To this end, various authors have proposed
alternative, more general definitions of a method's ``effective
degrees of freedom,'' or ``equivalent degrees of freedom'' (see
\cite{Buja1989} and references therein).

If the method is linear --- that is, if $\hat{\boldsymbol{y}} =
\boldsymbol{H}\boldsymbol{y}$ for some  ``hat
matrix'' $\boldsymbol{H}$ that is not a function of $\boldsymbol{y}$
--- then the trace of $\boldsymbol{H}$ serves as a natural generalization.
For linear regression $\bs H$ is a $p$-dimensional
projection, so $\tr(\boldsymbol{H}) = p$, coinciding with the original definition.
Intuitively, when $\boldsymbol{H}$ is not a projection,
$\tr(\boldsymbol{H})$  accumulates fractional degrees of
freedom for directions of $\bs{y}$ that are shrunk, but not
entirely eliminated, in computing $\hat{\bs{y}}$.

For nonlinear methods, further generalization
is necessary.  The most popular definition,
due to \cite{Efron1986} and given in
Equation~\eqref{df_epe_rss}, defines DF in terms of the optimism of
RSS as an estimate of test error, and applies to any fitting
method.

Measuring or estimating optimism is a worthy goal in and of itself.
But to justify our intuition that the DF offers a
consistent way to quantify model complexity, a bare requirement
is that the DF be monotone in model complexity when considering a
fixed method.

The term ``model complexity'' is itself rather metaphorical when
describing arbitrary fitting algorithms, but has a concrete meaning
for methods that minimize RSS subject to the fit $\hat{\bs{y}}$
belonging to a closed constraint set $\mathcal{M}$ (a ``model'').
Commonly, some tuning parameter $\gamma$ selects one of a nested set
of models:
\begin{equation}
  \hat{\bs{y}}^{(\gamma)} = \argmin_{\bs{z}\in\mathcal{M}_\gamma}
  \|\bs y - \bs{z}\|_2^2,\;\;\;\text{ with } \mathcal M_{\gamma_1}
  \subseteq \mathcal M_{\gamma_2}\subseteq \mathbb R^n
  \text{ if } \gamma_1 \leq \gamma_2
\end{equation}
Examples include the Lasso \citep{Tibshirani1996} and ridge regression
\citep{Hoerl1962} in their constraint formulation, as well as best
subsets regression (BSR).  The model $\mathcal M_k$ for BSR with $k$
variables is a union of $k$-dimensional subspaces.

\begin{figure}[ht!]\centering
\includegraphics[scale = 0.35]{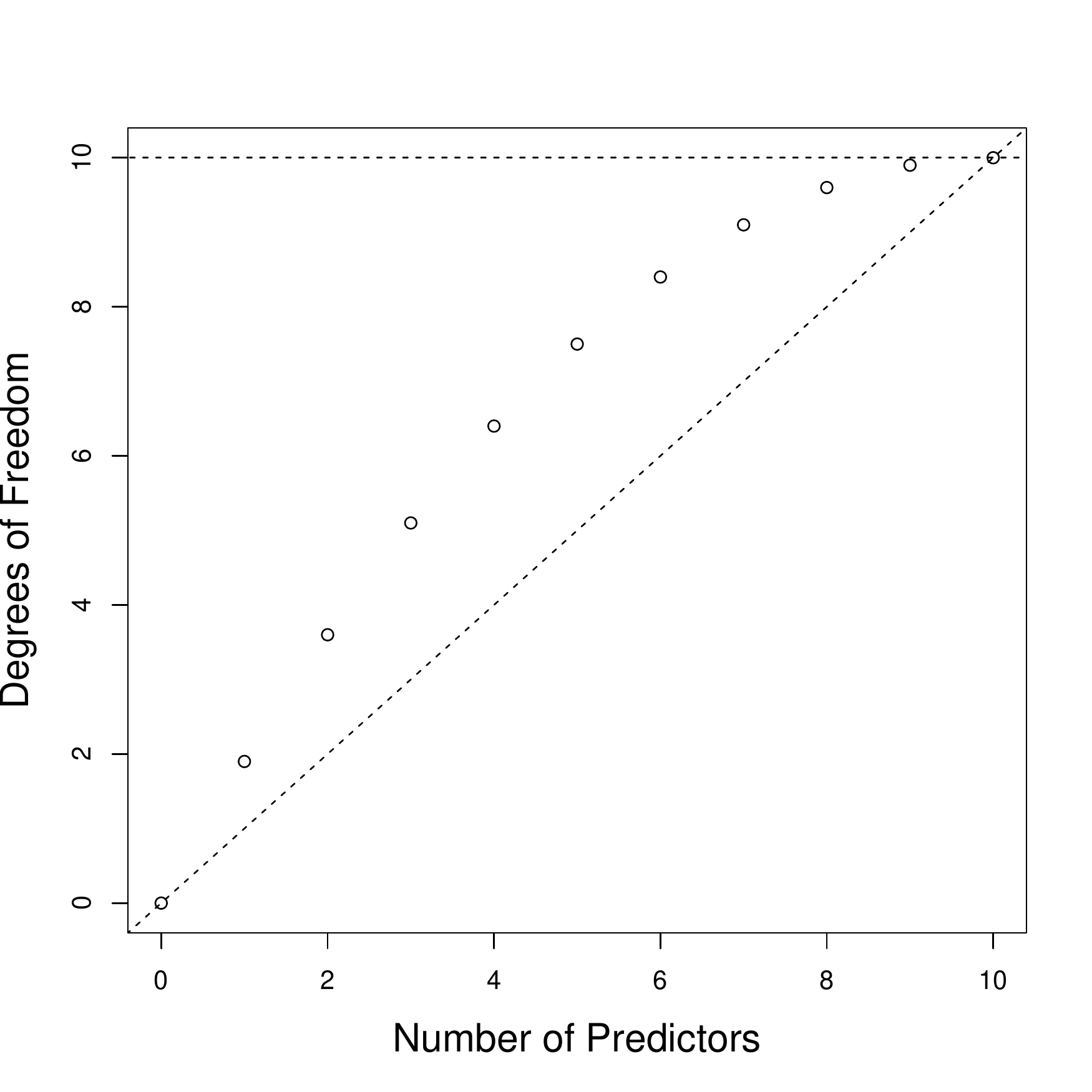}
\caption{This plot follows the usual intuition for degrees of freedom
  for best subset regression when the full model has 10 predictors.}
\label{dfFake}
\end{figure}

Because larger models ``fit harder'' to the data, one naturally
expects DF to be monotone with respect to model inclusion.
For example, one might imagine a plot of DF versus $k$ for BSR to
look like Figure~\ref{dfFake} when the full model has $p=10$
predictors: monotone and sandwiched between $k$ and $p$.

However, as we have already seen in Figure~\ref{fig:dfmap},
monotonicity is far from guaranteed even in very simple examples, and
in particular the DF ``ceiling'' at $p$ can be broken. Surprisingly,
monotonicity can even break down for methods projecting onto convex
sets, including ridge regression and the Lasso (although the DF cannot
exceed the dimension of the convex set). The non-monotonicity of
DF for such convex methods was discovered independently by
\cite{Kaufman2013}, who give a thorough account. Among other
results, they prove that the degrees of freedom of projection onto any
convex set must always be smaller than the dimension of that set. 

To minimize overlap with \cite{Kaufman2013}, we focus our attention instead on non-convex fitting procedures such as BSR. In contrast to the results of \cite{Kaufman2013} for the convex case, we show that projection onto any closed non-convex set can have arbitrarily large degrees of freedom, regardless of the dimensions of $\mathcal{M}$ and $\bs{y}$.

\section{Preliminaries}
\label{Prelim}
We consider
fitting techniques with some tuning parameter (discrete or continuous)
that can be used to vary the model from less to more constrained.  In BSR, the tuning
parameter $k$ determines how many predictor variables are retained in
the model, and we will denote BSR with tuning parameter $k$ as
BSR$_k$.
For a general fitting technique FIT, we will use the notation FIT$_k$
for FIT with tuning parameter $k$ and
$\hat{\boldsymbol{y}}^{{\scriptscriptstyle (FIT_k)}}$ for the
fitted response produced by FIT$_k$.

As mentioned in the introduction, a general formula for DF can be
motivated by the following relationship between expected prediction
error (EPE) and RSS for ordinary least
squares (OLS) \citep{Mallows1973}:
\begin{equation}
\text{EPE} = \mathbb{E}[\text{RSS}] + 2\sigma^2 p.
\end{equation}
Analogously, once a fitting technique (and tuning parameter) FIT$_k$ is
chosen for fixed data $\boldsymbol{y}$, DF is
defined by the following identity:
\begin{equation}
\label{df_epe_rss}
\mathbb{E}\left[\sum^n_{i = 1} (y^*_i - \hat{y}^{{\scriptscriptstyle
      (FIT_k)}}_i)^2\right] = \mathbb{E}\left[\sum^n_{i = 1} (y_i -
  \hat{y}^{{\scriptscriptstyle (FIT_k)}}_i)^2\right] + 2 \sigma^2
\cdot \text{DF} (\boldsymbol{\mu}, \sigma^2, \text{FIT}_k),
\end{equation}
where $\sigma^2$ is the variance of the $\varepsilon_i$ (which we assume
exists), and $y^*_i$ is a new independent copy of $y_i$ with mean
$\boldsymbol{\mu}_i$. Thus DF is defined as a measure of the optimism of
RSS. This definition in turn leads to a simple closed form
expression for DF under very general conditions, as shown by the
following theorem.

\begin{1}[\cite{Efron1986}]
\label{lemma1}
For $i \in \{1, \dots, n\}$, let $y_i = \mu_i + \varepsilon_i$, where
the $\mu_i$ are nonrandom and the $\varepsilon_i$ have mean zero and
finite variance.  Let $\hat{y}_i, \; i \in \{1, \dots, n\}$ denote
estimates of $\mu_i$ from some fitting technique based on a fixed
realization of the $y_i$, and let $y^*_i, \; i \in \{1, \dots, n\}$ be
independent of and identically distributed as the $y_i$.  Then
\begin{equation}
\mathbb{E}\left[\sum^n_{i = 1} (y^*_i - \hat{y}_i)^2\right] - \mathbb{E}\left[\sum^n_{i = 1} (y_i - \hat{y}_i)^2\right] = 2 \sum^n_{i = 1} \cov(y_i, \hat{y}_i)
\end{equation}
\begin{proof}
For $i \in \{1, \dots, n\}$,
\begin{equation}
\label{df1prf1}
\begin{split}
\mathbb{E}\big[(y^*_i - \hat{y}_i)^2\big] & = \mathbb{E}\big[(y^*_i - \mu_i + \mu_i - \hat{y}_i)^2\big] \\
& = \mathbb{E}\big[(y^*_i - \mu_i)^2\big] + 2\mathbb{E}\big[(y^*_i - \mu_i)(\mu_i - \hat{y}_i)\big] + \mathbb{E}\big[(\mu_i - \hat{y}_i)^2\big] \\
& = \var(\varepsilon_i) + \mathbb{E}\big[(\mu_i - \hat{y}_i)^2\big],
\end{split}
\end{equation}
where the middle term in the second line equals zero because $y^*_i$ is independent of all the $y_j$ and thus also of $\hat{y}_i$, and because $\mathbb{E}\big[y^*_i - \mu_i\big] = 0$. Furthermore,
\begin{equation}
\label{df1prf2}
\begin{split}
\mathbb{E}\big[(y_i - \hat{y}_i)^2\big] & = \mathbb{E}\big[(y_i - \mu_i + \mu_i - \hat{y}_i)^2\big] \\
& = \mathbb{E}\big[(y_i - \mu_i)^2\big] + 2\mathbb{E}\big[(y_i - \mu_i)(\mu_i - \hat{y}_i)\big] + \mathbb{E}\big[(\mu_i - \hat{y}_i)^2\big] \\
& = \var(\varepsilon_i) + \mathbb{E}\big[(\mu_i - \hat{y}_i)^2\big] - 2 \cov(y_i, \hat{y}_i).
\end{split}
\end{equation}
Finally, subtracting Equation~\eqref{df1prf2} from Equation~\eqref{df1prf1} and summing over ${i \in \{1, \dots, n\}}$ gives the desired result.
\end{proof}
\end{1}

\begin{2}
For i.i.d. errors with finite variance $\sigma^2$, Theorem~\ref{lemma1} implies that,
\begin{equation}
\label{dfcov}
\text{DF} (\boldsymbol{\mu}, \sigma^2, \text{FIT}_k) = \frac{1}{\sigma^2} \tr\big(\cov(\boldsymbol{y}, \hat{\boldsymbol{y}}^{{\scriptscriptstyle (FIT_k)}})\big) = \frac{1}{\sigma^2} \sum^n_{i = 1} \cov\big(y_i, \hat{y}^{{\scriptscriptstyle (FIT_k)}}_i\big).
\end{equation}
Note also that when FIT$_k$ is a linear fitting method with hat matrix
$\boldsymbol{H}$, Equation~\eqref{dfcov} reduces to
\begin{equation}
\label{dftrace}
\text{DF} (\boldsymbol{\mu}, \sigma^2, \text{FIT}_k) = \tr(\boldsymbol{H}).
\end{equation}
\end{2}


\section{Additional Examples}
\label{MotEx}
For each of the following examples, a model-fitting technique, mean
vector, and noise process are chosen. Then the DF is estimated by
Monte Carlo simulation. The details of this estimation process, along with all R
code used, are provided in the appendix.

\subsection{Best Subset Regression Example}
\label{example1}
Our first motivating example is meant to mimic a realistic application.
This example is a linear model with $n = 50$ observations on $p = 15$
variables and a standard Gaussian noise process.  The design matrix
$\boldsymbol{X}$ is a $n \times p$ matrix of \emph{i.i.d.} standard
Gaussian noise.  The mean vector $\boldsymbol{\mu}$ (or equivalently,
the coefficient vector $\boldsymbol{\beta}$) is generated by initially
setting the coefficient vector to the vector of ones, and then
standardizing $\boldsymbol{X}\boldsymbol{\beta}$ to have mean zero and
standard deviation seven.  We generated a few $(\bs X,
\bs \beta)$ pairs before we discovered one with substantially
non-monotonic DF, but then measured the DF for that $(\bs X,
\bs\beta)$ to high precision via Monte Carlo.
\begin{figure}[ht!]
\centering
\subfigure{\includegraphics[scale = 0.34]{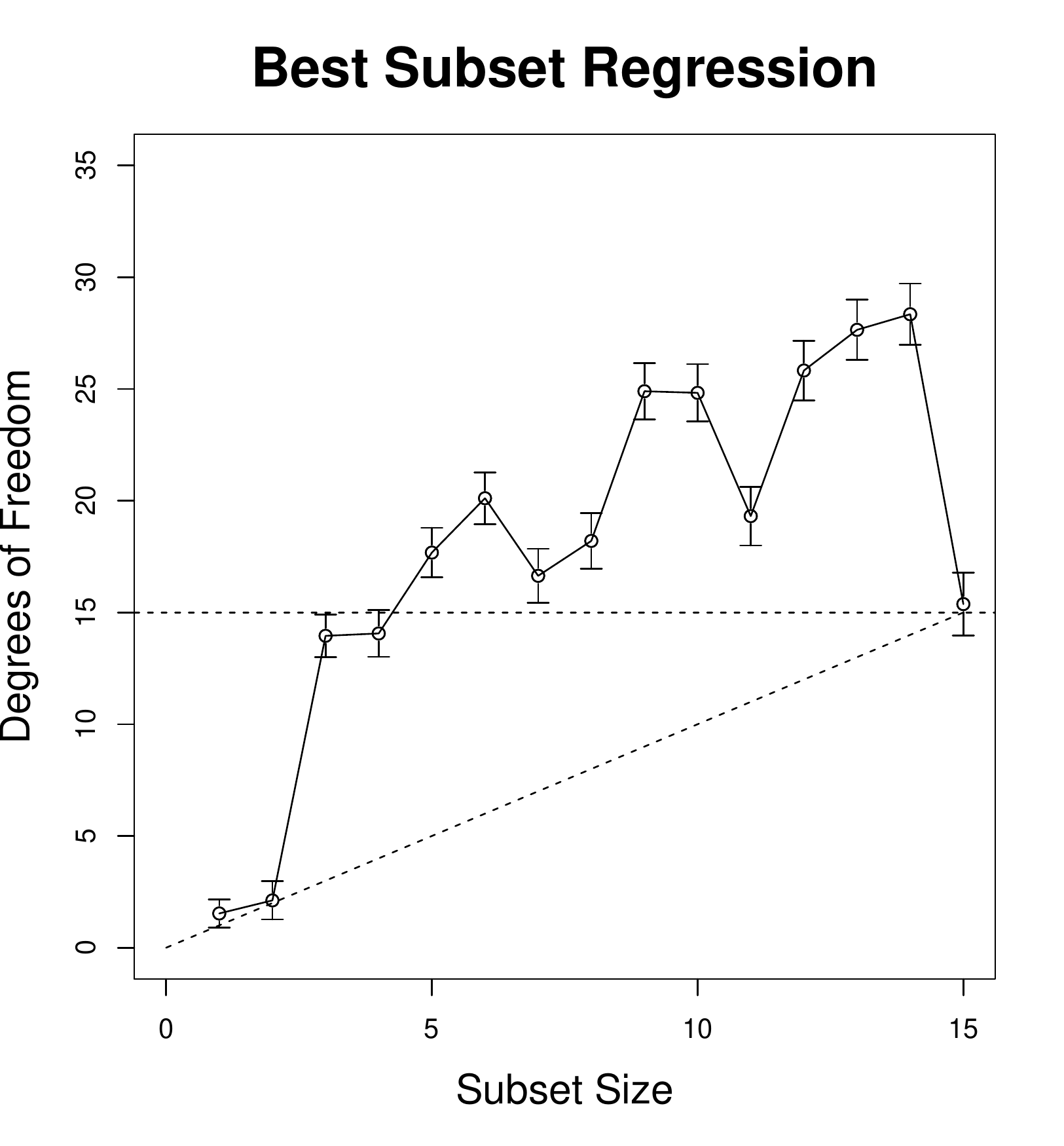}}
\subfigure{\includegraphics[scale = 0.34]{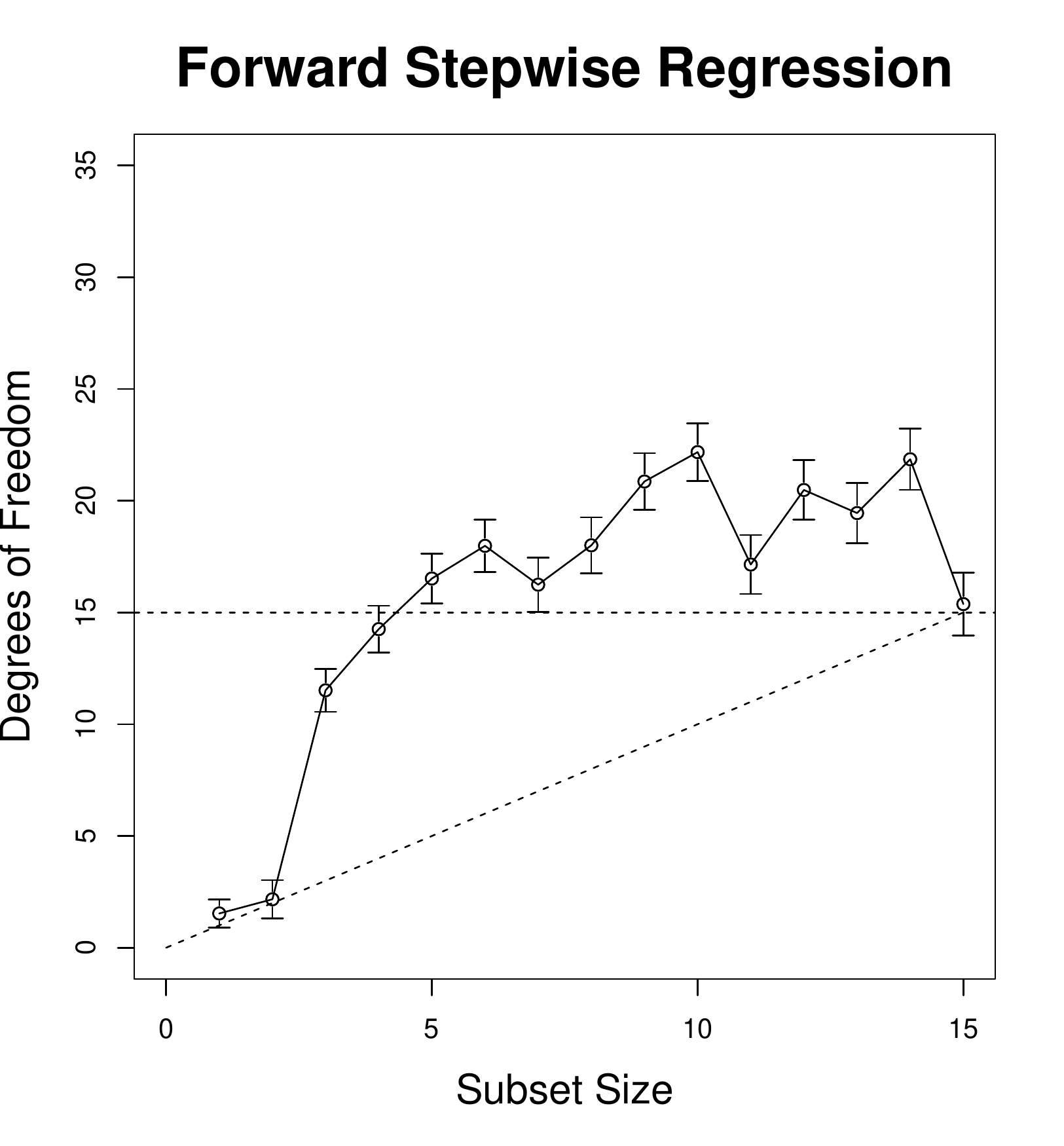}}
\caption{Monte Carlo estimated degrees of freedom versus subset size in the best
  subset regression (left) and forward stepwise regression (right)
  examples.  Estimates are shown plus or minus two (Monte Carlo)
  standard errors. Dashed lines show the constant ambient dimension
  ($p$) and the increasing subset size, for reference ($k$).}
\label{BSRFSRex}
\end{figure}

The left plot of Figure~\ref{BSRFSRex} shows the plot of Monte Carlo
estimated DF versus subset size. The right plot of
Figure~\ref{BSRFSRex} shows the same plot for forward stepwise
regression (FSR) applied to the same data. Although FSR isn't quite a
constrained least-squares fitting method, it is a popular method whose
complexity is clearly increasing in $k$. Both plots show that the DF for
a number of subset sizes is greater than 15, and we know from standard
OLS theory that the DF for subset size 15 is exactly 15 in both
cases.

\subsection{Unbounded Degrees of Freedom Example}
\label{example3}
\begin{figure}[ht!]
\centering
\includegraphics[scale = 0.4]{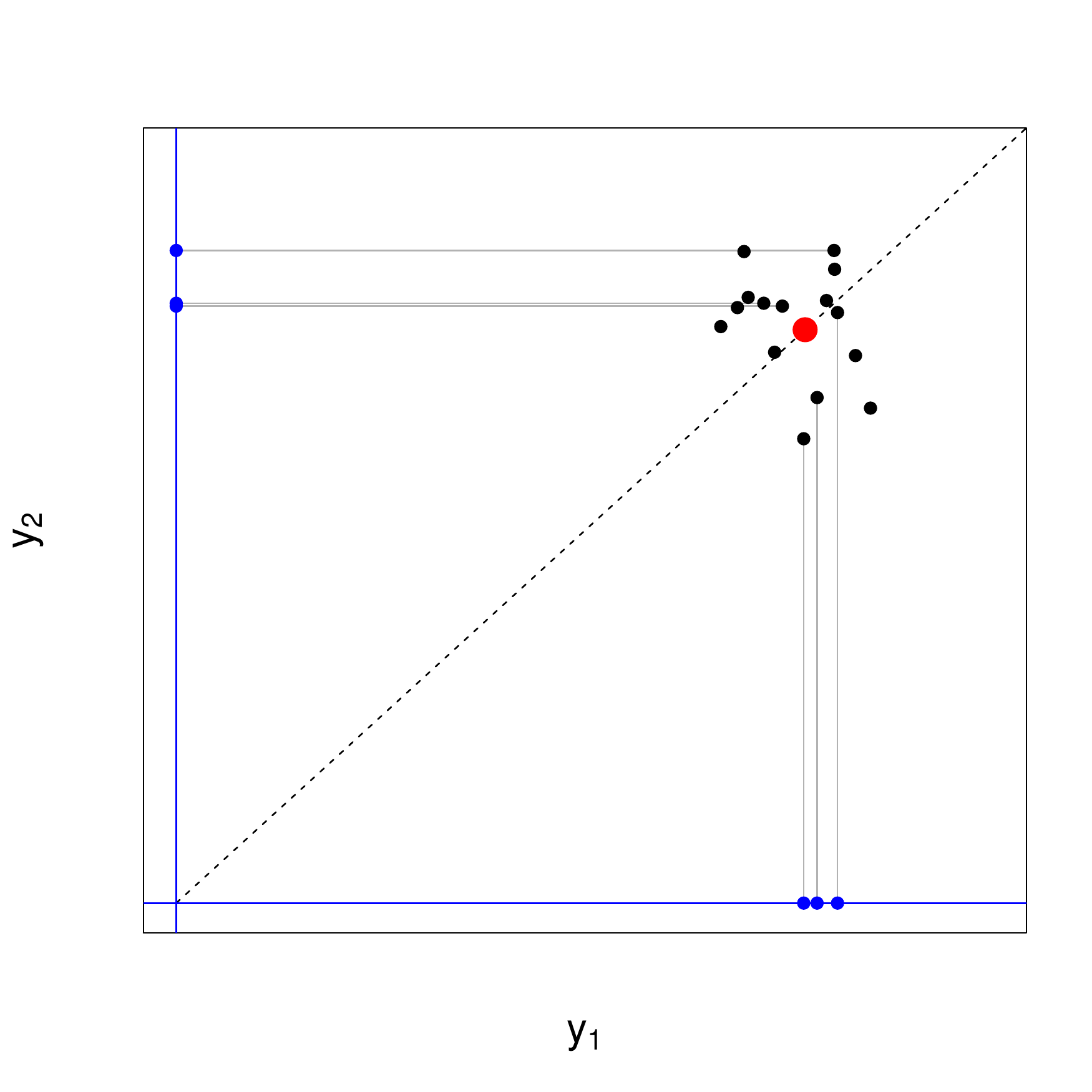}
\caption{Sketch of the unbounded DF example (not to scale with the $A$
  value used in the text). The true mean vector is
  in red, the constraint region in blue, and some data points are shown in
  black. The grey lines connect some of the data points to their
  projections onto the constraint set, in blue. Note that for ${i \in
  \{1, 2\}}$, $\hat{\boldsymbol{y}}_i$ is either 0 or approximately
  $A$ depending on small changes in $\boldsymbol{y}$.}
\label{PathEx}
\end{figure}

Next we return to the motivating example from Section~\ref{Intro}. The setup is again a linear model, but with
$n = p = 2$. The design matrix $\boldsymbol{X} = A\cdot
\boldsymbol{I}$, where $A$ is a scalar and $\boldsymbol{I}$ is the ($2
\times 2$) identity matrix. The coefficient vector is just $(1, 1)^T$
making the mean vector $(A, A)^T$, and the noise is \emph{i.i.d.}
standard Gaussian. Considering BSR$_1$, with high probability for
large $A$ (when $\bs{y}$ falls in the positive quadrant), it is clear
that the best univariate model just chooses the larger of the two response
variables for each realization. Simulating $10^5$ times with $A =
10^4$, we get a Monte Carlo estimate of the DF of BSR$_1$ to be about
5630, with a standard error of about 26.  Playing with the value of
$A$ reveals that for large $A$, DF is approximately linearly
increasing in $A$.  This suggests that not only is the DF greater than
$n = 2$, but can be made unbounded without changing $n$ or the
structure of $\boldsymbol{X}$. Figure~\ref{PathEx} shows what is going
on for a few points (and a smaller value of $A$ for visual
clarity). The variance of the $y_i$ (the black points) is far smaller
than that of the $\hat{y}_i$ (the projections of the black dots onto
the blue constraint set). Therefore, since the correlation between the
$y_i$ and the $\hat{y}_i$ is around 0.5, $\sum_{i=1}^n \cov(y_i,
\hat{y}_i)$ is also much larger than the variance of the $y_i$, and
the large DF can be infered from Equation~\eqref{dfcov}.

In this toy example, we can actually prove that the DF diverges as $A
\rightarrow \infty$ using Equation~\eqref{df_epe_rss},
\begin{equation}
\begin{split}
\frac{1}{A}\text{DF} ((A,A)^T, 1,
\text{BSR}_1) & = \frac{1}{2} \mathbb{E}\left[\frac{1}{A}\sum^2_{i =
    1} \Big((y^*_i - \hat{y}^{{\scriptscriptstyle (BSR_1)}}_i)^2 -
  (y_i - \hat{y}^{{\scriptscriptstyle (BSR_1)}}_i)^2\Big)\right] \\
& = \frac{1}{2} \mathbb{E}\left[\frac{1}{A}\sum^2_{i = 1} \Big((y^*_i
  - \hat{y}^{{\scriptscriptstyle (BSR_1)}}_i)^2 - (y_i -
  \hat{y}^{{\scriptscriptstyle (BSR_1)}}_i)^2\Big)\mathbb{I}_{\bs{y}
    \in Q_1}\right] \\
& \quad + \frac{1}{2} \mathbb{E}\left[\frac{1}{A}\sum^2_{i = 1} \Big((y^*_i
  - \hat{y}^{{\scriptscriptstyle (BSR_1)}}_i)^2 - (y_i -
  \hat{y}^{{\scriptscriptstyle (BSR_1)}}_i)^2\Big)\mathbb{I}_{\bs{y}
    \notin Q_1}\right] \\
& = \frac{1}{2} \mathbb{E}\left[\frac{1}{A}\Big(A^2 +
  2A\varepsilon_1^* + (\varepsilon_1^*)^2 + \big(\varepsilon_2^* -
  \min (\varepsilon_1, \varepsilon_2)\big)^2 \right. \\
& \quad \qquad \qquad \left.- A^2 - 2A\min (\varepsilon_1,
  \varepsilon_2) - \big(\min (\varepsilon_1,
  \varepsilon_2)\big)^2\Big)\mathbb{I}_{\bs{y} \in Q_1}\right] \\
& \quad + o(1) \\
& \stackrel{A \rightarrow \infty}{\longrightarrow} \frac{1}{2}\mathbb{E}\left[2\varepsilon_1^* -
  2\min(\varepsilon_1, \varepsilon_2) \right] =
\mathbb{E}\left[\max(\varepsilon_1, \varepsilon_2)\right], \\
\end{split}
\end{equation}
where $Q_1$ is the first quadrant of $\mathbb{R}^2$, $\mathbb{I}_S$
is the indicator function on the set $S$, and $\varepsilon_1^*,
\varepsilon_2^*$ are noise realizations independent of one another
and of $\bs{y}$. The $o(1)$ term comes from the the fact that $\frac{1}{A}\sum^2_{i = 1} \Big((y^*_i
  - \hat{y}^{{\scriptscriptstyle (BSR_1)}}_i)^2 - (y_i -
  \hat{y}^{{\scriptscriptstyle (BSR_1)}}_i)^2\Big)$ is $O_p(A)$ while
  $\mathbb{P}(\bs{y} \notin Q)$ shrinks exponentially fast in $A$
  (as it is a Gaussian tail probability). The convergence in the last
  line follows by the Dominated Convergence Theorem applied to the
  first term in the preceding line. Note that
  $\mathbb{E}\left[\max(\varepsilon_1, \varepsilon_2)\right] \approx
  0.5642$, in good agreement with the simulated result.

\section{Geometric Picture}
\label{geometric}
The explanation for the non-monotonicity of DF can be best understood
by thinking about regression problems as constrained
optimizations. We will see that the reversal in DF monotonicity is
caused by the interplay between the shape of the coefficient
constraint set and that of the contours of the objective function.

\begin{figure}[ht!]
\centering
\includegraphics[scale=0.4]{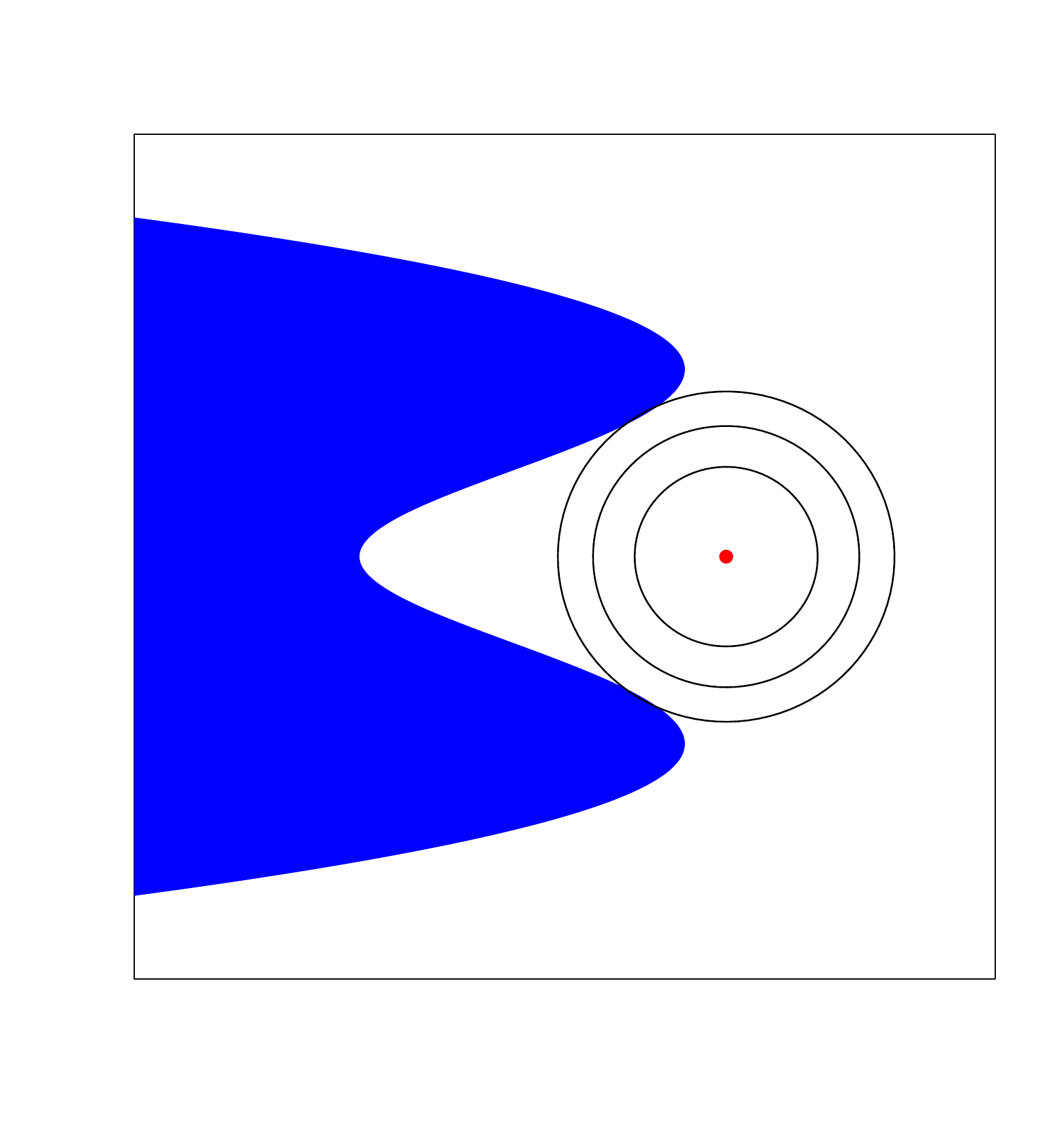}
\caption{Sketch of geometric picture for a regression problem with
  $n=2$ and non-convex constraint sets. The blue area is the
  constraint set, the red point the true mean vector, and the black
  circles the contours of the least squares objective function. Note
  how the spherical contour spans the gap of
  the divot where it meets the constraint set.}
\label{fig:nonconvex}
\end{figure}

BSR has a non-convex constraint set for $k < p$, and
Subsection~\ref{example3} gives some intuition for why the DF can be
much greater than the model dimension. This intuition can be generalized to any non-convex
constraint set, as follows. Place the true mean at a point with
non-unique projection onto the constraint set, see
Figure~\ref{fig:nonconvex} for a generic sketch. Such a point must exist
by the Motzkin-Bunt Theorem \citep{Bunt1934, Motzkin1935,
  Kritikos1938}. Note the constraint set for
$\hat{\boldsymbol{y}} = \boldsymbol{X}\hat{\boldsymbol{\beta}}$ is
just an affine transformation of the constraint set for
$\hat{\boldsymbol{\beta}}$, and thus a non-convex
$\hat{\boldsymbol{\beta}}$ constraint is equivalent to a non-convex
$\hat{\boldsymbol{y}}$ constraint. Then the fit depends sensitively on
the noise process, even when the noise is very small, since
$\boldsymbol{y}$ is projected onto multiple well-separated sections of the
constraint set. Thus as the magnitude of the noise, $\sigma$, goes
to zero, the variance of $\hat{\boldsymbol{y}}$ remains roughly
constant. Equation~\eqref{dfcov} then tells us that DF can be made
arbitrarily large, as it will be roughly proportional to
$\sigma^{-1}$. We formalize this intuition in the following theorem.

\begin{5}
\label{nonConvexThm}
For a fitting technique FIT that minimizes squared error subject to a
non-convex, closed constraint $\hat{\boldsymbol{\mu}} \in \mathcal{M}
\subset \mathbb{R}^n$, consider the model,
\[ \boldsymbol{y} = \boldsymbol{\mu} + \sigma \boldsymbol{\varepsilon},
\quad \varepsilon_i \iid F \text{ for } i \in \{1, \dots, n\}, \]
where $F$ is a mean-zero distribution with finite variance
supported on an open neighborhood of 0. Then there exists some $\boldsymbol{\mu}^*$ such that
$\text{DF}(\boldsymbol{\mu}^*, \sigma^2, \text{FIT}_k) \rightarrow \infty$
as $\sigma^2 \rightarrow 0$.
\begin{proof}
We leave the proof for the appendix.
\end{proof}
\end{5}

\section{Discussion}
\label{discussion}

The common intuition
that ``effective'' or ``equivalent'' degrees of
freedom serves a consistent and interpretable measure of model
complexity merits some degree of skepticism.  Our results and examples, combined with those of \cite{Kaufman2013}, demonstrate that for many widely-used convex and non-convex fitting techniques, the DF can be
non-monotone with respect to
model nesting. In the non-convex case, the DF can exceed the dimension of the model space by an
arbitrarily large amount.
This phenomenon is not restricted
to pathological examples or edge cases, but can arise in run-of-the-mill
datasets as well. Finally, we presented a geometric interpretation of
this phenomenon in terms of irregularities in
the model constraints and/or objective function contours.

In light of the above, we find the term
``(effective) degrees of freedom'' to be somewhat misleading, as it is
suggestive of a quantity corresponding to model ``size''. It is also
misleading to consider DF as a measure of overfitting, or how flexibly
the model is conforming to the data, since a model always fits at
least as strongly as a strict submodel to a given dataset.
By definition, the DF of \cite{Efron1983} does measure
optimism of in-sample error as an estimate of out-of-sample error, but we
should not be too quick to assume that our intuition about linear
models carries over exactly.

\bibliography{library}

\newcommand{\noop}[1]{}
\begin{thebibliography}{}

\bibitem[Buja et~al., 1989]{Buja1989}
Buja, A., Hastie, T., and Tibshirani, R. (1989).
\newblock {Linear smoothers and additive models}.
\newblock {\em The Annals of Statistics}, 17(2):453--510.

\bibitem[Bunt, 1934]{Bunt1934}
Bunt, L. N.~H. (1934).
\newblock {\em {Bijdrage tot de theorie der convexe puntverzamelingen}}.
\newblock PhD thesis, Rijks-Universiteit, Groningen.

\bibitem[Efron, 1983]{Efron1983}
Efron, B. (1983).
\newblock {Estimating the error rate of a prediction rule: improvement on
  cross-validation}.
\newblock {\em Journal of the American Statistical Association},
  78(382):316--331.

\bibitem[Efron, 1986]{Efron1986}
Efron, B. (1986).
\newblock {How Biased Is the Apparent Error Rate of a Prediction Rule?}
\newblock {\em Journal of the American Statistical Association},
  81(394):461--470.

\bibitem[Hoerl, 1962]{Hoerl1962}
Hoerl, A.~E. (1962).
\newblock {Application of Ridge Analysis to Regression Problems}.
\newblock {\em Chemical Engineering Progress}, 58:54--59.

\bibitem[Kaufman and Rosset, 2014]{Kaufman2013}
Kaufman, S. and Rosset, S. (2014).
\newblock {When Does More Regularization Imply Fewer Degrees of Freedom?
  Sufficient Conditions and Counter Examples from Lasso and Ridge Regression}.
\newblock {\em Biometrika (to appear)}.

\bibitem[Kritikos, 1938]{Kritikos1938}
Kritikos, M.~N. (1938).
\newblock {Sur quelques propri\'{e}t\'{e}s des ensembles convexes}.
\newblock {\em Bulletin Math\'{e}matique de la Soci\'{e}t\'{e} Roumaine des
  Sciences}, 40:87--92.

\bibitem[Mallows, 1973]{Mallows1973}
Mallows, C.~L. (1973).
\newblock {Some Comments on CP}.
\newblock {\em Technometrics}, 15(4):661--675.

\bibitem[Motzkin, 1935]{Motzkin1935}
Motzkin, T. (1935).
\newblock {Sur quelques propri\'{e}t\'{e}s caract\'{e}ristiques des ensembles
  convexes}.
\newblock {\em Atti Accad. Naz. Lincei Rend. Cl. Sci. Fis. Mat. Natur},
  21:562--567.

\bibitem[Tibshirani, 1996]{Tibshirani1996}
Tibshirani, R. (1996).
\newblock {Regression shrinkage and selection via the lasso}.
\newblock {\em Journal of the Royal Statistical Society. Series B},
  58(1):267--288.

\end{thebibliography}
\bibliographystyle{apalike}

\section{Appendix A}
\begin{proof}[Proof of Theorem~\ref{nonConvexThm}]
The proof relies heavily on the fact that for every non-convex set
$\cal M$ in Euclidean space there is at least one point whose
projection onto $\cal M$ is not unique (e.g. the red dot in
Figure~\ref{fig:nonconvex}). This fact was proved independently in
\cite{Bunt1934}, \cite{Motzkin1935}, and \cite{Kritikos1938}. A
schematic for this proof in two dimensions is provided in
Figure~\ref{fig:proofpic}.
\begin{figure}[ht!]
\centering
\includegraphics[scale = 0.6]{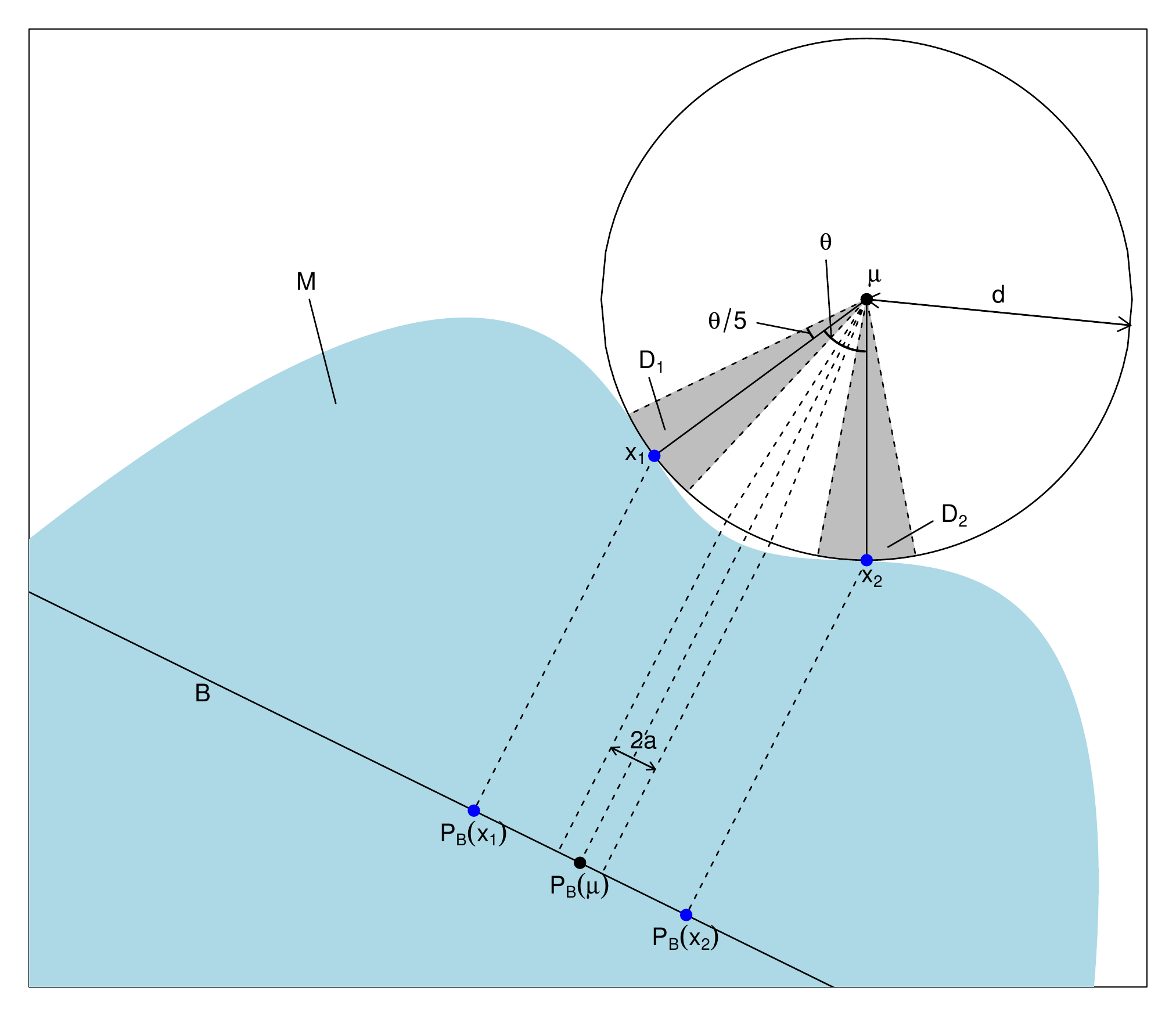}
\caption{Schematic for the proof of Theorem~\ref{nonConvexThm}, in two
  dimensions.}
\label{fig:proofpic}
\end{figure}
Let $\bs{\mu}$ be a point with non-unique projection onto the
non-convex set $\mathcal{M}$ and let $\bs{x}_1$ and $\bs{x}_2$ be two
distinct projections of $\bs{\mu}$ onto $\mathcal{M}$. Let $d =
||\bs{\mu} - \bs{x}_1||_2 = ||\bs{\mu} - \bs{x}_2||_2$ be the
Euclidean distance between $\bs{\mu}$ and $\mathcal{M}$, and $\theta =
\cos^{-1}(\frac{(\bs{x}_1 - \bs{\mu}) \cdot (\bs{x}_2 -
  \bs{\mu})}{|\bs{x}_1 - \bs{\mu}||\bs{x}_2 - \bs{\mu}|})$ be the
angle between $\bs{x}_1$ and $\bs{x}_2$, taken as vectors from
$\bs{\mu}$. Define the set $\mathcal{D}_1 = \{\bs{v} \in \mathbb{R}^n
| \cos^{-1}(\frac{(\bs{x}_1 - \bs{\mu}) \cdot (\bs{v} -
  \bs{\mu})}{|\bs{x}_1 - \bs{\mu}||\bs{v} - \bs{\mu}|}) <
\frac{\theta}{5}, ||\bs{v} - \bs{\mu}||_2 < d \}$, and
$\mathcal{D}_2$ analogously for $\bs{x}_2$. Let $\mathcal{B}$ be a
one-dimensional affine subspace that is both parallel to the line
connecting $\bs{x}_1$ and $\bs{x}_2$, and contained in the hyperplane
defined by $\bs{\mu}$, $\bs{x}_1$, and $\bs{x}_2$. Denoting the
projection operator onto $\mathcal{B}$ by $P_{\mathcal{B}}$, let $z =
||P_{\mathcal{B}}\bs{y} - P_{\mathcal{B}}\bs{\mu}||_2 / \sigma$, and
$\tilde{y} = ||P_{\mathcal{B}}\hat{\bs{y}} -
P_{\mathcal{B}}\bs{\mu}||_2$. Let $a = d \cos(\frac{\pi - \theta /
  5}{2})$. We now have,
\begin{equation*}
\begin{split}
\tr(\cov(\bs{y}, \hat{\bs{y}})) & \ge \cov(P_{\mathcal{B}}\bs{y},
P_{\mathcal{B}}\hat{\bs{y}}) \\
& = \mathbb{E}[\sigma z \tilde{y}] \\
& = \mathbb{E}[\sigma z \tilde{y} 1_{y \in \mathcal{D}_1 \cup
  \mathcal{D}_2}] + \mathbb{E}[\sigma z \tilde{y} 1_{y \notin \mathcal{D}_1 \cup
  \mathcal{D}_2}] \\
& \ge \sigma \mathbb{E}[z \tilde{y} 1_{y \in \mathcal{D}_1}] + \sigma
\mathbb{E}[z \tilde{y} 1_{y \in \mathcal{D}_2}] \\
& \ge a \sigma \big (\mathbb{E}[z 1_{y \in \mathcal{D}_2}] -
\mathbb{E}[z 1_{y \in \mathcal{D}_1}] \big ), \\
& = 2 a \sigma \mathbb{E}[z 1_{y \in \mathcal{D}_2}]. \\
\end{split}
\end{equation*}
The first inequality follows from the translation and rotation
invariance of the trace of a covariance matrix, and from the
positivity of the diagonal entries of the covariance matrix for the
case of projection fitting methods. For the second inequality, note that $\mathbb{E}[\sigma z \tilde{y} 1_{y \notin \mathcal{D}_1 \cup \mathcal{D}_2}]\geq 0$,
again because of the positivity of the DF of projection methods
(applied to the same model with a noise process that has support on
$\mathcal{D}_1$ and $\mathcal{D}_2$ removed). The third
inequality follows from considering the projections of $\mathcal{D}_1$ and
$\mathcal{D}_2$ onto $\mathcal{M}$ and then onto $\mathcal{B}$, and
noting that the two (double) projections must be separated by at least
a distance of $2a$.

Defining $\mathcal{F}_1 = \{\bs{v} \in \mathbb{R}^n
| \cos^{-1}(\frac{(\bs{x}_1 - \bs{\mu}) \cdot (\bs{v} -
  \bs{\mu})}{|\bs{x}_1 - \bs{\mu}||\bs{v} - \bs{\mu}|}) <
\frac{\theta}{5} \}$ and $\mathcal{F}_2$ analogously for $\bs{x}_2$,
note that $\mathbb{P}(\bs{y} \in \mathcal{F}_1 \setminus
\mathcal{D}_1) = \mathbb{P}(\bs{y} \in \mathcal{F}_2 \setminus
\mathcal{D}_2) \rightarrow 0$ as $\sigma^2 \rightarrow 0$. Thus,
\begin{equation*}
\begin{split}
\tr(\cov(\bs{y}, \hat{\bs{y}})) & \ge 2 a \sigma \big (\mathbb{E}[z 1_{y
  \in \mathcal{F}_2}] + o(\sigma) \big ). \\
\end{split}
\end{equation*}
Neither $z$ nor the event $y \in \mathcal{F}_2$ depend on
$\sigma$, so define the constant $b = 2 a \mathbb{E}[z 1_{y \in
  \mathcal{F}_2}] > 0$ which is independent of $\sigma$. Thus we have
shown that,
\begin{equation*}
\begin{split}
\text{DF}(\bs{\mu}^*, \sigma^2, \text{FIT}_k) & = \frac{1}{\sigma^2} \tr(\cov(\bs{y}, \hat{\bs{y}})) \\
& \ge \frac{b + o(\sigma)}{\sigma} \\
& \rightarrow \infty
\end{split}
\end{equation*}
as $\sigma^2 \rightarrow 0$.
\end{proof}

\section{Appendix B}
In Subsections~\ref{example1} and \ref{example3}, DF is estimated by
computing an ubiased estimator of DF for each simulated noise
realization. This unbiased estimator for DF can be obtained from Equation~\eqref{dfcov} by exploiting the linearity of the expectation and trace operators,
\begin{equation}
\begin{split}
\label{dfest}
\text{DF} (\boldsymbol{\mu}, \sigma^2, \text{FIT}_k) & = \frac{1}{\sigma^2} \tr(\cov(\boldsymbol{y}, \hat{\boldsymbol{y}}^{{\scriptscriptstyle (FIT_k)}})) \\
& = \frac{1}{\sigma^2} \mathbb{E}\big [(\boldsymbol{y} - \boldsymbol{\mu})^T (\hat{\boldsymbol{y}}^{{\scriptscriptstyle (FIT_k)}} - \mathbb{E}[\hat{\boldsymbol{y}}^{{\scriptscriptstyle (FIT_k)}}])\big ] \\
& = \frac{1}{\sigma^2} \mathbb{E}\big [\boldsymbol{\varepsilon}^T \hat{\boldsymbol{y}}^{{\scriptscriptstyle (FIT_k)}}\big ],
\end{split}
\end{equation}
where the last inequality follows because
$\mathbb{E}[\boldsymbol{\varepsilon}] = \boldsymbol{0}$. Thus the true
DF is estimated by $\boldsymbol{\varepsilon}^T
\hat{\boldsymbol{y}}^{{\scriptscriptstyle (FIT_k)}}$ averaged over
many simulations. Since this estimate of DF is an average of \emph{i.i.d.}
random variables, its standard deviation can be estimated by the
emipirical standard deviation of the $\boldsymbol{\varepsilon}^T
\hat{\boldsymbol{y}}^{{\scriptscriptstyle (FIT_k)}}$ divided by the
square root of the number of simulations.

The following is the code for the BSR simulation in
Subsection~\ref{example1}.  Note that although the seed is preset,
this is only done to generate a suitable design matrix, not to
cherry-pick the simulated noise processes. The simulation results
remain the same if the seed is reset after assigning the value to
\lstinline{x}.
\begin{lstlisting}
library(leaps)
library(gplots)
n = 50; p = 15; B = 5000
set.seed(1113)
x = matrix(rnorm(n * p), n, p)
mu = drop(scale((x %*% rep(1, 15))))
snr = 7
mu = mu * snr
dfmat = matrix(0, B, p)
for(j in 1:B){
	y = rnorm(n) + mu
	#temp = regsubsets(x, y, nbest = 1, nvmax = p)
        temp = regsubsets(x, y, nbest = 1, nvmax = p, intercept = FALSE)
	for(i in 1:p){
		jcoef = coef(temp, id = i)
		#xnames = names(jcoef)[-1]
                xnames = names(jcoef)
		which = match(xnames, letters[1:p])
		#yhat = cbind(1, x[ , which]) %*% jcoef
                if(i == 1){
                    yhat = matrix(x[ , which], n, 1) %*% jcoef
                } else{
                    yhat = x[ , which] %*% jcoef
                }
		dfmat[j, i] = sum((y - mu) * yhat)
	}
}
#df = apply(dfmat, 2, mean) - 1
df = apply(dfmat, 2, mean)
error = sqrt(apply(dfmat, 2, var) / B)
save(df, error, p, file = "BSR.rdata")
plotCI(1:p, df, 2 * error, ylim = c(0, 35), xlim = c(0, p), gap = 0.4, xlab = "Subset Size", ylab = "Degrees of Freedom", cex.lab = 1.5, type = "o")
abline(h = 15, lty = 2)
lines(c(0, 15), c(0, 15), lty = 2)
title("Best Subset Regression", cex.main = 2)
\end{lstlisting}

The code for the FSR simulation in Subsection~\ref{example1} is almost
identical.
\begin{lstlisting}
library(leaps)
library(gplots)
n = 50; p = 15; B = 5000
set.seed(1113)
x = matrix(rnorm(n * p), n, p)
mu = drop(scale((x %*% rep(1, 15))))
snr = 7
mu = mu * snr
dfmat = matrix(0, B, p)
for(j in 1:B){
	y = rnorm(n) + mu
	#temp = regsubsets(x, y, nbest = 1, nvmax = p, method = "forward")
        temp = regsubsets(x, y, nbest = 1, nvmax = p, method = "forward", intercept = FALSE)
	for(i in 1:p){
		jcoef = coef(temp, id = i)
		#xnames = names(jcoef)[-1]
                xnames = names(jcoef)
		which = match(xnames, letters[1:p])
		#yhat = cbind(1, x[ , which]) %*% jcoef
                if(i == 1){
                    yhat = matrix(x[ , which], n, 1) %*% jcoef
                } else{
                    yhat = x[ , which] %*% jcoef
                }
                dfmat[j, i] = sum((y - mu) * yhat)
	}
}
#df = apply(dfmat, 2, mean) - 1
df = apply(dfmat, 2, mean)
error = sqrt(apply(dfmat, 2, var) / B)
save(df, error, p, file = "FSR.rdata")
plotCI(1:p, df, 2 * error, ylim = c(0, 35), xlim = c(0, p), gap = 0.4, xlab = "Subset Size", ylab = "Degrees of Freedom", cex.lab = 1.5, type = "o")
abline(h = 15, lty = 2)
lines(c(0, 15), c(0, 15), lty = 2)
title("Forward Stepwise Regression", cex.main = 2)
\end{lstlisting}

Finally, the code for Subsection~\ref{example3} is different but short.  We can imagine taking $A \rightarrow \infty$.
\begin{lstlisting}
# A is some big number
A = 10000
X = A * data.frame(diag(2))
B = 1E5
y = A + matrix(rnorm(2 * B), ncol = 2)

# best univariate (no-intercept) model just picks the larger y_i
yhat.bs = t(apply(y, MARGIN = 1, FUN = function(yy) yy * (yy > mean(yy))))

dfvec = drop(((y - A) * yhat.bs) %*% c(1, 1))
df = mean(dfvec)
error = sqrt(var(dfvec) / B)
c(df, error)
\end{lstlisting}

\end{document}